# Criteria of Science, Cosmology, and Lessons of History[*]

Helge Kragh[†]

**Abstract**  Perhaps more than any other science, cosmology exemplifies the inevitable contact between science and philosophy, including the problem of the demarcation criteria that distinguish science from non-science. Although modern physical cosmology is undoubtedly scientific, it is not obvious why it has this status, and nor is it obvious that all branches of theoretical cosmology satisfy ordinarily assumed criteria for science. While testability is generally admitted as an indispensable criterion for a theory being scientific, there is no agreement among cosmologists what testability means, more precisely. For example, should testability be taken to imply falsifiability in the sense of Popper? I discuss this and related questions by referring to two episodes of controversy in the history of modern cosmology, the debate over the steady state theory in the 1950s and the recent debate concerned with the anthropic multiverse. In addition, I draw attention to the use of historical analogies in cosmological and other scientific arguments, suggesting that such use is often misuse or otherwise based on distortions of the history of science.

## 1. Introduction

Ever since the age of Galileo, at the beginning of the Scientific Revolution, science has expanded in both breadth and depth, conquering one area after the other. The development of the scientific enterprise has not occurred at a uniform growth rate, of course, but it has nonetheless been remarkably successful, progressing cognitively as well as socially and institutionally. Today, some 400 years after Galileo first demonstrated the inadequacy of the

---

Aristotelian cosmos and the advantages of the Copernican alternative, we may wonder if there are any limits at all to scientific inquiry. Will science at some future stage enable us to understand everything? Is scientific explanation limitless? These are big questions and not really the topic of this essay, but I shall nevertheless introduce it by some general reflections on the limits of science, divided in four points.

(i) When it comes to the question of the limits of science, it is useful to distinguish between *knowledge* and *explanation*. After all, we may have scientific knowledge about things, even understand them on a phenomenological or instrumentalist level, and yet be unable to provide them with an explanation. Indeed, the history of science is one long series of temporary disharmonies between phenomenal and explanatory knowledge. Early radioactivity is one example of an unexplained phenomenon that nonetheless was investigated in great detail and with great success. Another example is superconductivity, which was discovered in 1911 but only explained on a microphysical basis with the BCS (Bardeen-Cooper-Schrieffer) theory dating from 1957.

(ii) The question of scientific explanation obviously depends on our chosen criteria for what constitutes an acceptable *explanation*.[1] These criteria are not provided by nature, but by the scientific community. With an appropriate change of the criteria scientists may be able to explain phenomena that previously seemed inexplicable. This point is particularly well illustrated by the anthropic principle, which provides explanations for a variety of phenomena – from the neutron-proton mass difference to the age of the universe – that cannot be explained on the basis of standard physics and cosmology. But are anthropic explanations proper explanations at all? As well known, this is a matter of considerable debate and a main reason why the anthropic principle is controversial.[2]

---

[1] The philosophical literature on scientific and other explanations is extensive. Relevant works include R. Nozick, *Philosophical Explanations*, Harvard University Press, Cambridge MA 1981, P. Achinstein, *The Nature of Explanation*, Oxford University Press, Oxford 1983, and J. Cornwell, ed., *Explanation: Styles of Explanation in Science*, Oxford University Press, Oxford 2004.

(iii) Implicitly or explicitly, the question of the limits of science refers to the problem of the *domain* of science, that is, the territory of reality to which science applies. Are there phenomena or concepts that lie outside the realm of science, or can science legitimately be applied to *all* aspects of reality? According to hard-core reductionists the latter is the case. Thus, Frank Tipler is by his own admission an "uncompromising reductionist," implying that "everything, including human beings, can be completely described by physics."[3] Generally, within the tradition of positivism the tendency has been to define reality as just those phenomena or concepts that are accessible to scientific analysis.

However, it is possible that the world that can be observed in principle (and hence be subject to scientific analysis) is only part of a larger non-physical world to which we have no empirical access and which therefore transcends the domain of science as ordinarily understood. For example, this is what has been argued within a non-theistic context by Milton Munitz, a distinguished philosopher of cosmological thought. According to him, there is a dimension of existence, which he calls "Boundless Existence," that transcends the existence of the physical universe. This Boundless Existence is not in space and time, it has no structure, and it can only be characterized – if characterized at all – in negative terms. "Boundless Existence," Munitz says, "is so totally unique … that all similarities with anything in our ordinary experience must fall short and be inadequate."[4]

---

[2] See, for example, R. J. Deltete, *What Does the Anthropic Principle Explain?*, "Perspectives on Science" 1993, no. 1, pp. 285-305.

[3] F. J. Tipler, *The Physics of Immortality: Modern Cosmology, God, and the Renaissance of the Dead*, Doubleday, New York 1994, p. 352.

(iv) There are questions of a conceptual nature about which we do not even know whether they are meaningful or not – or, if they are meaningful, whether they belong to the domain of science. To indicate the type of these questions, a brief reference to two problems may suffice. First, there is the much discussed question of realized or actual infinities, of whether or not there can be an infinite number of objects in the universe. The problem has become an issue in the standard inflationary model of the flat universe, but it was also discussed in relation to the earlier steady state model according to which space was infinite and uniformly populated with matter. While many modern cosmologists are perfectly happy with actual infinities, others deny their scientific legitimacy and consider the question to be metaphysical.[5] The point is that we do not really know whether or not it makes scientific sense. It makes mathematical and philosophical sense, but will it ever be answered scientifically?

If infinity is one of those frightening concepts on the border between physics and metaphysics, so is the concept of *nothingness* or absolute void. This is another speculation with a rich and fascinating history that recently has become relevant to science, not least after the discovery of the dark energy revealed by the acceleration of the cosmic expansion. Dark energy is generally identified with the vacuum energy density as given by the cosmological constant. However, whether or not this turns out to be true, the modern quantum vacuum is entirely different from absolute nothingness. As far as I can see, there cannot possibly be a scientific answer to what nothingness is, and yet it does not therefore follow that the concept is meaningless.[6] Such a conclusion presupposes a rather narrow positivistic perspective.

---

[4] M. K. Munitz, *Cosmic Understanding: Philosophy and Science of the Universe*, Princeton University Press, Princeton 1986, p. 235. See also M. K. Munitz, *The Question of Reality*, Princeton University Press, Princeton 1990.

[5] G. F. R. Ellis, U. Kirchner, W. R. Stoeger, *Multiverses and Physical Cosmology*, "Monthly Notices of the Royal Astronomical Society" 2004, no. 347, pp. 921-936. On the disturbing infinities appearing in steady state cosmology, see R. Schlegel, *The Problem of Infinite Matter in Steady-State Cosmology*, "Philosophy of Science" 1965, no. 32, pp. 21-31.

In this essay I look at a fundamental question in the philosophy of science, namely, the defining criteria of what constitutes scientific activity from a cognitive point of view. Another and largely equivalent version of this question is the demarcation problem, that is, how to distinguish between science and non- or pseudoscience. Why is astronomy recognized as a science, when astrology and gastronomy are not? However, I shall not deal with these questions in a general and abstract way, but instead illustrate some of them by means of a couple of examples from the more recent history of cosmology. I focus on two cases, the one being the controversy related to the steady state theory in the 1950s and the other the still ongoing controversy over the anthropic multiverse. Although separated in time by half a century, in some respects they are surprisingly similar and suited for comparison.

One remarkable feature shared by the two cases is the role played by philosophical considerations among the scientists themselves – philosophy *in* rather than *of* science.[7] The history of cosmology, and the history of science more generally, demonstrates that on the fundamental level philosophy is not extraneous to science but part and parcel of it. I suggest that Freeman Dyson was quite wrong when he stated, in a rare mood of positivism, that, "philosophy is nothing but empty words if it is not capable of being tested by experiments."[8] As will become clear, the views of science associated with Karl Popper's critical philosophy played an important role in both controversies. For this reason, I deal particularly with these views and Popper's emphasis on testability and falsifiability as defining criteria for science also in the area of physical cosmology. In the last section I offer some reflections on the use and misuse of historical analogies in the evaluation of scientific theories, a problem that turned up in both of the cosmological controversies.

---

[6] A useful overview is presented in R. Sorensen, *Nothingness*, "Stanford Encyclopedia of Philosophy" 2003, http://plato.stanford.edu/entries/nothingness. See also B. Rundle, *Why there is Something Rather than Nothing*, Oxford University Press, Oxford 2004. For the history of the concepts of vacuum and nothingness, see H. Genz, *Nothingness: The Science of Empty Space*, Basic Books, New York 1999.

## 2. Testability in the physical sciences

Few modern philosophers of science believe that science can be defined methodologically in any simple way and, at the same time, reflect the actual historical course of science.[9] There is no generally accepted, more or less invariant formulation that encapsulates the essence of science and its rich variation. All the same, there are undoubtedly *some* criteria of science and theory choice that almost all scientists agree upon and have accepted for at least two centuries. Thomas Kuhn suggested five such standard criteria of evaluation, which he took to be (1) accuracy; (2) consistency, internal as well as external; (3) broadness in scope; (4) simplicity; (5) fruitfulness.[10] Although Kuhn did not mention testability as a separate criterion, it was part of the first one, according to which there must be "consequences deducible from a theory [that] should be in demonstrated agreement with the results of existing experiments and observations." Kuhn did not specifically refer to predictions, except that he included them under the notion of "fruitfulness."

---

[7] On the concept of "philosophy in science" and some of the problems related to it, see M. Heller, *How Is Philosophy in Science Possible?*, [in:] *Philosophy in Science*, eds. B. Brozek, J. Maczka, W. P. Grygiel, Copernicus Center Press, Krakow 2011, pp. 13-24.

[8] F. Dyson, *Infinite in All Directions*, Perennial, New York 2004, p. 96. A balanced argument for the value of philosophy in cosmological research is given in E. McMullin, *Is Philosophy Relevant to Cosmology?*, "American Philosophical Quarterly" 1981, no. 18, pp. 177-189.

[9] This section relies on material discussed more fully in a paper on *Testability and Epistemic Shifts in Modern Cosmology* submitted to "Studies in History and Philosophy of Modern Physics".

Most philosophers of science, including Kuhn himself, are aware, that the mentioned criteria may contradict each other in concrete situations and that a relative weighing may therefore be needed. But then the system cannot fully or uniquely determine an evaluation in a concrete case. In the context of modern cosmology Kuhn's criteria have been discussed by George Ellis, who points out that although they are all desirable they are not of equal relevance and may even lead to conflicts, that is, to opposing conclusions with regard to theory choice.[11] Still, Ellis (and most other cosmologists) finds the first of Kuhn's criteria to be the one that in particular characterizes a scientific theory and demarcates it from other theories. In short, empirical testability is more than just one criterion out of many. Nearly all scientists consider this epistemic value an indispensable criterion for a theory being scientific: a theory which is cut off from confrontation with empirical data just does not belong to the realm of science.

---

[10] T. S. Kuhn, *The Essential Tension: Selected Studies in Scientific Tradition and Change*, University of Chicago Press, Chicago 1977, pp. 321-322.

[11] G. F. R. Ellis (2007), *Issues in the Philosophy of Cosmology*, [in:] *Philosophy of Physics*, eds. J. Butterfield, J. Earman, North-Holland, Amsterdam 2007, pp. 1183-1286.

As an example, consider Einstein, who in the period from about 1905 to 1925 moved from a cautious empiricist position *à la* Mach to an almost full-blown rationalism. In his Herbert Spencer lecture of 1933 he famously stated that "we can discover by means of pure mathematical considerations the concepts and the laws …, which furnish they key to the understanding of natural phenomena. … In a certain sense, therefore, I hold it true that pure thought can grasp reality, as the ancients dreamed."[12] But in between these two expressions of his rationalist credo, there was the no less important sentence: "Experience remains, of course, the sole criterion of the physical utility of a mathematical construction." As late as 1950, commenting on his latest attempt at a generalized theory of gravitation, he readily admitted that "Experience alone can decide on truth."[13] According to Einstein, while in the creative or constructive phase of a scientific theory empirical considerations might be wholly absent, such considerations were at the very heart of the context of justification.

While testability is universally admitted as a necessary (but not, of course, sufficient) condition for a theory being scientific, in practice the concept can be interpreted in ways that are so different that the consensus may tend to become rhetorical only and of little practical consequence. The following list of interpretive questions is not complete, but it gives an idea of

---

[12] A. Einstein, *Ideas and Opinions*, Three Rivers Press, New York 1982. On Einstein's philosophy of science, see, for example, J. Shelton, *The Role of Observation and Simplicity in Einstein's Epistemology*, "Studies in History and Philosophy of science" 1988, no. 19, pp. 103-118, and J. D. Norton, *"Nature is the Realization of the Simplest Conceivable Mathematical Ideas": Einstein and the Canon of Mathematical Simplicity*, "Studies in History and Philosophy of Modern Physics" 2000, no. 31, pp. 135-170.
[13] A. Einstein, *On the Generalized Theory of Gravitation*, "Scientific American" 1950, no. 182:4, pp. 13-17, on p. 17.

what physicists sometimes disagree about when it comes to testing of theories:

1. Actual testability (with present instrument technologies or those of a foreseeable future) is obviously preferable. But should it be required that a theory is actually testable, or will testability in principle – perhaps in the form of a thought experiment – suffice?
2. Should a theory result in precise and directly testable predictions, or will indirect testability do? For example, if a fundamental theory $T$ results in several successfully confirmed predictions $P_1$, $P_2$, …, $P_n$, can prediction $P_{n+1}$ be considered to have passed a test even if it is not actually tested?[14]
3. Will a real test have to be empirical, by comparing consequences of the theory with experiments or observations, or do mathematical consistency checks also count as sufficient (theoretical) tests?
4. Another kind of non-empirical testing is by way of thought experiments or arguments of the *reductio ad absurdum* type that played an important role in the controversy over the steady state theory. A cosmological model may lead to consequences that are either contradictory or unacceptably bizarre. How should such arguments enter the overall evaluation picture?

---

[14] It is sometimes argued that there are reasons to believe in untestable predictions if they follow from a well-established theory with empirical success. On this account the existence of other universes is "tested" by the successfully tested background theories, in this case quantum mechanics and inflation theory. See, for example, M. Tegmark, *The Mathematical Universe*, "Foundations of Physics" 2008, no. 38, pp. 101-150. On a different note, string theorists have suggested that the theory of superstrings has passed an empirical test because it includes gravitation without being designed to do so. E. Witten, *Magic, Mystery, and Matrix*, "Notices of the AMS" 1998, no. 45, 1124-1129.

5. At what time in the development of a theory or research programme can one reasonably demand testability? Even if a theory is not presently testable, perhaps it will be so in a future version, such as there are many examples of in the history of science.
6. How should (lack of) testability be weighed in relation to (lack of) other epistemic desiderata? E.g., is an easily testable theory with a poor explanatory record always to be preferred over a non-testable theory with great explanatory power? Or what if the testable theory is overly complicated, and the non-testable one is mathematically unique and a paragon of simplicity?
7. Should predictions of novel phenomena be counted as more important than pre- or postdictions of already known phenomena? This is a question on which philosophers are divided and where the historical evidence is ambiguous.

## 3. A historical case: The steady state theory

The steady state theory of the universe, proposed by Fred Hoyle, Hermann Bondi and Thomas Gold in 1948, aroused a great deal of philosophical interest, in part because of the theory's controversial claim of continual creation of matter and more generally because of its appeal to philosophy and methods of science. For example, Bondi and Gold argued that the new steady state theory was preferable from a methodological point of view, as it was simpler, more direct, and more predictive than the cosmological theories based on general relativity. The latter class of theories, they said, was "utterly unsatisfactory" since it covered a whole spectrum of theories that could only be confronted with the observed universe if supplied with more or less arbitrary assumptions and parameters: "In general relativity a very wide range of models is available and the comparisons [between theory and observation] merely attempt to find which of these models fits the facts best. The number of free parameters is so much larger than the number of observational points that a fit certainly exists and not even all the parameters can be fixed."[15] Relativistic cosmology sorely lacked the deductive character of the steady state theory, which uniquely led to a number of predictions, such as the mean density of matter, the curvature of space, and the average age of galaxies. According to Bondi and Gold, the predictions were crucially based on what they called the perfect cosmological principle (PCP), namely, the postulate that there is neither a privileged place nor a privileged time in the universe. Thus, the PCP is a temporal extension of the ordinary cosmological principle (CP).

Whether in the Bondi-Gold or the Hoyle version, the steady state theory was critically discussed by many philosophers and philosophically minded astronomers and physicists.[16] To the first category belonged Adolf Grünbaum, Mario Bunge, Milton Munitz, Norwood Russell Hanson, and Rom Harré, and to the latter Herbert Dingle, Gerald Whitrow, William McCrea, and William Bonnor. We witness in this discussion an instructive case of philosophy in science, an unusual dialogue between professional philosophers and the spontaneous philosophy of practicing scientists.

---

[15] H. Bondi and T. Gold, *The Steady-State Theory of the Expanding Universe*, "Monthly Notices of the Royal Astronomical Society" 1948, no. 108, pp. 252-270, on p. 269 and p. 262.

[16] On the philosophical foundation of steady state cosmology and the discussion of its scientific status, see Y. Balashov, *Uniformitarianism in Cosmology: Background and Philosophical Implications of the Steady-State Theory*, "Studies in History and Philosophy of Science" 1994, no. 25, pp. 933-958, and H. Kragh, *Cosmology and Controversy: The Historical Development of Two Theories of the Universe*, Princeton University Press, Princeton 1996.

Much of the methodological discussion in the 1950s and 1960s focused on the criteria on which to judge the scientific nature of the steady state theory, or of cosmology in general. To give just a couple of examples, Dingle found the cosmological principle – whether in its original CP or the "perfect" PCP form – to be plainly unscientific because it was a priori and hence in principle inviolable.[17] According to Bunge and some other critics, the steady state theory was nothing but "science-fiction cosmology" because it rested on the ad hoc assumption of continual creation of matter.[18] On the other hand, and contrary to the later multiverse controversy, testability was not at the heart of the discussion. Both parties accepted that a cosmological theory must be observationally testable, but they rated this epistemic value somewhat differently and did not draw the same conclusions from it.

In 1954 Bondi and Whitrow engaged in an interesting public debate concerning the scientific status of physical cosmology. Whitrow, stressing the unique domain of cosmology, argued that it was not truly scientific and probably never would be so. It would remain, he thought, a borderland subject between science and philosophy. Bondi, on the other hand, suggested that the hallmark of science was falsifiability of theories and that on this criterion cosmology was indeed a science. "Every advocate of any [cosmological] theory will always be found to stress especially the supposedly excellent agreement between the forecasts of his theory and the sparse observational results," he admitted. And yet,

---

[17] H. Dingle, *Cosmology and Science*, [in:] *The Universe*, eds. G. Piel et al., Simon and Schuster, New York 1956, pp. 131-138. The misguided claim that the cosmological principle is a priori has more recently been made by the German philosopher Kurt Hübner, according to whom cosmological models rest on a priori constructions that are essentially independent of observations. K. Hübner, *Critique of Scientific Reason*, University of Chicago Press, Chicago 1985, pp. 150-152.

[18] M. Bunge, *Cosmology and Magic*, "The Monist" 1962, no. 47, pp. 116-141.

> The acceptance of the possibility of experimental and observational disproof of any theory is as universal and undisputed in cosmology as in any other science, and, though the possibility of logical disproof is not denied in cosmology, it is not denied in any other science either. By this test, the cardinal test of any science, modern cosmology must be regarded as a science. … I consider universal acceptance of the possibility of experimental disproof of any claim an absolute test of what constitutes a science.[19]

Although not mentioning Karl Popper by name, Bondi was clearly defending a main methodological point in Popperian philosophy which he much admired. Whitrow, who was also well acquainted with Popper's views, did not disagree, although he warned that falsifiability should not be considered a final and absolute criterion: "The important role of disproof in science, which has been so cogently argued by K. R. Popper, is intimately related to the self-correcting tendency of science and this, in my view, is another aspect of the pursuit of unanimity."[20]

Although Popperian criteria of science played a considerable role during the cosmological controversy, and were highlighted by the steady state proponents in particular, they were rarely an issue of dispute. By and large, criteria of a Popperian kind were accepted also by many cosmologists favouring an evolving universe governed by the laws of general relativity. One of them was the British astronomer George McVittie, who was strongly opposed to the steady state theory and other theories he suspected were based on a priori principles. He described the philosophical foundation of the Bondi-Gold theory as "Karl Popper's dictum that a scientific theory can never be proved to be true but, instead, that certain theories can be proved to be false by an appeal to observation." While he considered the dictum to be a "probably unimpeachable doctrine," he parodied Bondi's use of it. If one followed Bondi's vulgar version of Popper's philosophy, "we should be justified in inventing a theory of gravitation which would prove that the orbit of every planet was necessarily a circle. The theory would be most vulnerable to observation and could, indeed, be immediately shot down."[21]

---

[19] G. J. Whitrow, H. Bondi, *Is Physical Cosmology a Science?*, "British Journal for the Philosophy of Science" 1954, no. 4, pp. 271-283, on p. 279 and p. 282. For the Bondi-Whitrow discussion, see also H. Kragh, *Cosmology and Controversy*, op. cit., pp. 233-237.

[20] Whitrow, Bondi, *Is Physical Cosmology a Science?*, op. cit., p. 280.

## 4. A modern case: The anthropic multiverse

Like the earlier controversy over the steady state cosmological model, the present discussion of the multiverse hypothesis deals to a large extent with philosophical issues and the borderline between science and philosophy.[22] Both cases are about foundational issues that cannot be answered simply by observation and calculation. Among those issues are: Does the theory belong to science proper, or is it rather a philosophical speculation? If it disagrees with established standards of science, should these standards perhaps be changed? What are the basic criteria for deciding whether a theory is true or false? The discussion in 2008 between Bernard Carr and George Ellis concerning the multiverse, taking place in the journal *Astronomy &*

---

[21] G. C. McVittie, *Rationalism versus Empiricism in Cosmology*, "Science" 1961, no. 133, 1231-1236, on p. 1231. McVittie belonged to what he called the "observational school" in cosmology. See J.-M. Sánchez-Ron, *George McVittie, the Uncompromising Empiricist*, [in:] *The Universe of General Relativity*, eds. A. J. Kox, Jean Eisenstaedt, Birkhäuser, Boston 2005, pp. 189-222.

[22] The central source in the multiverse debate is *Universe or Multiverse*, ed. B. Carr, Cambridge University Press, Cambridge 2007. See also H. Kragh, *Higher Speculations: Grand Theories and Failed Revolutions in Physics and Cosmology*, Oxford University Press, Oxford 2011, where further references are given. More popular accounts of the multiverse (in one or more of its several versions) include L. Susskind, *The Cosmic Landscape: String Theory and the Illusion of Intelligent Design*, Little, Brown and Company, New York 2006, and A. Vilenkin, *Many Worlds in One: The Search for other Universes*, Hill and Wang, New York 2006.

*Geophysics*, can be seen as a modern analogue of the 1954 Bondi-Whitrow discussion about the scientific nature of physical cosmology.[23]

However, although the two cosmological controversies have enough in common to make a comparison meaningful, there are also some dissimilarities. As mentioned, in the case of the steady state theory there was a great deal of interest from the side of the philosophers, who were key players in the debate. Strangely, a corresponding interest is largely absent in the case of the multiverse debate, where the philosophically related questions are predominantly discussed by the physicists themselves. Another difference is that the overarching question of the multiverse hypothesis is whether or not it is testable by ordinary observational means. Does it result in predictions of such a kind that, should they turn out to be wrong, the hypothesis must be wrong as well? In this respect the cases of the steady state and the multiverse are quite different: whereas the first was eminently falsifiable – and was in fact falsified – the multiverse fares very badly in terms of falsifiability. As has often been pointed out, it explains a lot but predicts almost nothing.

The current discussion concerning the multiverse involves two major questions of obvious relevance to the philosophy of and in science:

> (i) Has cosmology become truly and exclusively scientific, in the sense that philosophical considerations no longer play a legitimate role? If so, has it achieved this remarkable status by changing the rules of science?

---

[23] B. Carr, G. F. R. Ellis, *Universe or Multiverse?*, "Astronomy & Geophysics" 2008, no. 49, pp. 2.29-2.37.

16(ii) Which people or groups have the "right" to define these rules of science and thus to decide whether or not a particularly theory discussed by the scientists is in fact scientific?

It is far from clear whether some of the recent developments, such as multiverse cosmology and aspects of so-called physical eschatology, belong primarily to science or philosophy. The idea of many universes, traditionally a subject of philosophical speculation, is now claimed to have been appropriated by the physical sciences. Is this yet another conquest of the ever-victorious physics, at the expense of philosophy? According to Max Tegmark, this is indeed the case. "The borderline between physics and philosophy has shifted quite dramatically in the last century," he comments. "Parallel universes are now absorbed by that moving boundary. It's included within physics rather than metaphysics."[24] However, sceptics disagree.

One problem with the multiverse hypothesis is that the excessive amount of universes seems to allow almost any physical state of affairs – if not in our universe, then in some other. This, together with the unobservability of the other universes, tends to make the multiverse unacceptable from Popperian-like points of view. According to Popper's philosophy, a scientific theory must be falsifiable and therefore set constraints to the results of possible observations: "Every 'good' scientific theory is a prohibition: it forbids certain things to happen," as he said in a lecture of 1953.[25] At least in some versions, multiverse cosmology suffers from an extreme lack of prohibitiveness.

Some physicists advocating the multiverse and anthropic reasoning have questioned whether there is any need for external norms of science of a philosophical nature, these norms being Popperian or something else. "If scientists need to change the borders of their own field of research," says the French cosmologist Aurélien Barrau, "it would be hard to justify a philosophical prescription preventing them from doing so."[26] Leonard Susskind, the leading advocate of the string-based landscape multiverse theory, agrees with Barrau that normative prescriptions are unnecessary and may even be harmful. He suggests that only the scientists themselves, or perhaps their scientific communities, can decide by means of their practices what is and what is not science: "It would be very foolish to throw away the right answer on the basis that it doesn't conform to some criteria for what is or isn't science."[27] Susskind is particularly dissatisfied with the falsification criterion and what he calls the "overzealous Popperism" advocated by the "Popperazi" following Popper's philosophy. "Throughout my long experience as a scientist," he says, "I have heard unfalsifiability hurled at so many important ideas that I am inclined to think that no idea can have great merit unless it has drawn this criticism. … Good scientific methodology is not an abstract set of rules dictated by philosophers."[28]

---

[24] Quoted in C. Seife, *Physics Enters the Twilight Zone*, "Science" 2004, no. 305, p. 465.
[25] K. R. Popper, *Conjectures and Refutations*, Routledge, New York 1963, p. 48.

It needs to be pointed out that the Barrau-Susskind argument is deeply problematic and hardly tenable. Not only is it circular reasoning to define science as what scientists do, it also presupposes that all scientists have roughly the same ideas of what constitutes science, which is definitely not the case. Not even within such a relatively small field as theoretical cosmology is there any consensus. Subjects that scientists find interesting and discuss at conferences or write articles about in peer-reviewed journals do not automatically belong to the realm of science. Moreover, it makes no sense to speak of a "right answer" without appealing, explicitly or implicitly, to some criteria of science. To conclude that a theory is either valid or invalid necessarily involves certain standards of scientific validity. These standards need not be part of a rigid philosophical system ("dictated by philosophers"), nor do they have to be explicitly formulated, but it is hard to see how they can be avoided. Nature herself does not provide us with the criteria for when an answer is right.

---

[26] A. Barrau, *Physics in the Universe*, "Cern Courier" 2007 (20 November, online edition).
[27] Quoted in G. Brumfiel, *Outrageous Fortune*, "Nature" 2006, no. 358, p. 363.
[28] L. Susskind, *The Cosmic Landscape, op. cit.*, pp. 193-195. See also H. Kragh, *Higher Speculations, op. cit.*, pp. 280-285.

**5. Karl Popper and modern cosmology**

As already indicated, Popper's philosophy of science has played, and continues to play, an important role in methodological debates concerning cosmology. According to a study by Benjamin Sovacool, astronomers and cosmologists often invoke Popper's ideas as a guide for constructing and evaluating theories, although they rarely reveal a deeper familiarity with these ideas.[29] The first time Popperianism entered the scene of cosmology was in the 1950s, in connection with the steady state theory and Bondi's explicit use of standards based on Popper's philosophy of science. In a discussion of modern cosmology from 1960, he summarized Popper's view as follows: "The purpose of a theory is to make forecasts that can be checked against observation and experiment. A scientific theory is one that it is in principle possible to disprove by empirical means. It is this supremacy of empirical disproof that distinguishes science from other human activities. … A scientific theory, to be useful, must be testable and vulnerable."[30]

---

[29] B. Sovacool, *Falsification and Demarcation in Astronomy and Cosmology*, "Bulletin of Science, Technology & Society" 2005, no. 25, pp. 53-62.

The leading theoretical physicist and cosmologist Lee Smolin is no less a "Popperazo" than Bondi was. As Bondi used Popper's philosophy to criticize the big bang theory, so Smolin uses it to dismiss most versions of multiverse cosmology. "According to Popper," he says, "a theory is falsifiable if one can derive from it unambiguous predictions for practical experiments, such that – were contrary results seen – at least one premise of the theory would have been proven not true. … Confirmation of a prediction of a theory does not show that the theory is true, but falsification of a prediction can show it is false."[31]

In regard of the considerable impact of Popper's thoughts, it is remarkable that physical cosmology is hardly mentioned at all in his main works. Yet a closer look reveals that cosmology does turn up in his books and papers, most explicitly in a lecture given in 1982 in Alpbach, Austria. Calling cosmology "the most philosophically important of all the sciences," at this occasion he praised the by then defunct Bondi-Gold-Hoyle theory as "a very fine and promising theory," not because it was true but because it was testable and had in fact been falsified. As a result of measurements based on methods of radio astronomy, "it seems to have been refuted in favour of the (older) big bang theory of expansion."[32] Popper did not mention the cosmic microwave background radiation or other evidence (such as the measured amount of helium in the universe) that had laid the steady state theory in the grave.

---

[30] H. Bondi, *The Steady-State Theory of the Universe*, [in:] *Rival Theories of Cosmology*, eds. H. Bondi et al., Oxford University Press, London 1960, pp. 12-21, on p. 12.
[31] L. Smolin, *Scientific Alternatives to the Anthropic Principle*, [in:] *Universe or Multiverse?*, ed. B. Carr, Cambridge University Press, Cambridge 2007, pp. 323-366, on pp. 323-324. Emphasis added. For Smolin as a self-declared "Popperazo," see L. Smolin, *The Trouble with Physics*, Penguin Books, London 2008, p. 369.
[32] K. R. Popper, *In Search of a Better World: Lectures and Essays from Thirty Years* 1994, Routledge, London, pp. 58-60. For a full account of Popper's view of cosmology and the impact of his philosophy on cosmologists, see H. Kragh, *"The Most*

Although references to Popper's philosophy of science often appear in modern cosmology, it is probably fair to say that few of the physicists and astronomers have actually read him. Most seem to rely on what they have been told or happen to know from the secondary literature. This results in discussions that are sometimes simplistic and based on misunderstandings. What cosmologists (and other scientists) discuss is most often naïve falsificationism rather than the sophisticated versions of authentic Popperianism.[33] Popper's views, including his awareness that falsifiability cannot stand alone as a demarcation criterion, were far from the caricatures one can sometimes meet in the science literature. It should be recalled that his philosophy was normative and that he did not claim that the associated standards reflected the actual practice of scientists. Moreover, he never held that falsifiability is a sufficient condition for a theory being scientific, but only that it is a necessary condition. Although somewhat ambiguous with regard to the relationship between his methodological rules and scientific practice, he admitted that strict falsifiability does not belong to the real world of science:

> In point of fact, no conclusive disproof of a theory can ever be produced; for it is always possible to say that the experimental results are not reliable, or that the discrepancies which are asserted to exist between the experimental results and the theory are only apparent and that they will

---

*Philosophically Important of All the Sciences": Karl Popper and Physical Cosmology*, http://philsci-archive.pitt.edu/id/eprint/9062, a version of which will appear in a forthcoming issue of *Perspectives on Science*.

[33] As pointed out in M. Heller, *Ultimate Explanations of the Universe*, Springer-Verlag, Berlin 2009, pp. 88-89.

> disappear with the advance of our understanding. … If you insist on strict proof (or strict disproof) in the empirical sciences, you will never benefit from experience, and never learn from it how wrong you are.[34]

Contrary to what many scientists believe, Popper did not assign any absolute value to the criterion of falsifiability and did not consider it a *definition* of science. He recognized that the distinction between metaphysics and science is often blurred. "What was a metaphysical idea yesterday can become a testable theory tomorrow," he wrote.[35] Far from elevating falsificationism to an inviolable principle, he suggested that it is itself fallible and that it may be rational to keep even an admittedly wrong theory alive for some time. Popper wrote as follows:

> There is a legitimate place for dogmatism, though a very limited place. He who gives up his theory too easily in the face of apparent refutations will never discover the possibilities inherent in his theory. *There is room in science for debate*: for attack and therefore also for defence. Only if we try to defend them can we learn all the different possibilities inherent in our theories. As always, science is conjecture. You have to conjecture when to stop defending a favourite theory, and when to try a new one.[36]

This is indeed a view far from the strict or naïve falsificationism often discussed by scientists either for or against Popper. It is a view closer to the one associated with philosophers of science such as Imre Lakatos and Thomas Kuhn.

---

[34] K. R. Popper, *The Logic of Scientific Discovery*, Basic Books, New York 1959, p. 50. In a note appended to the English edition, Popper remarked that "I have been constantly misinterpreted as upholding a criterion (and moreover one of *meaning* rather than of *demarcation*) based upon a doctrine of 'complete' or 'conclusive' falsifiability."

[35] K. R. Popper, *Replies to my Critics*, [in:] *The Philosophy of Karl Popper*, ed. P. A. Schilpp, Open Court Publishing House, La Salle, IL 1974, pp. 961-1200, on p. 981.

[36] K. R. Popper, *Replies to my Critics*, *op.cit.*, p. 984. Popper's emphasis.

## 6. The role of historical analogies

Just like scientists use methodological and other philosophical arguments in evaluating the value of a fundamental scientific theory, sometimes they use (or misuse) arguments relating to the history of science. The typical way of doing this is by supporting an argument of a philosophical kind by means of concrete historical cases in the form of exemplars. That is, history is used analogically. The standard formula is this: Since, in a certain historical case, the epistemic value *x* proved successful, a modern theory should preferably incorporate *x*; or, conversely, if values of the kind *y* have proved a blind alley in the past, they should be avoided in a modern theory. The values or prescriptions *x* and *y* will usually be those associated with either well known successes or failures in the history of science. Often it is enough to associate them with the great authorities of the past.

Historical analogy arguments of this kind are quite common in controversies and in discussions of theories of a foundational nature. Einstein often relied on historical exemplars when he wanted to illustrate and give support to his favourite deductivist methodology of science, such as he did in the semi-popular book *The Evolution of Physics*.[37] During the cosmological controversy in the 1950s, some physicists and astronomers used Galileo's supposed empiricism as a weapon against what they considered rationalistic and a priori tendencies in the steady state theory. McVittie associated this theory with Aristotle's dogmatic world system (!) and the empirical cosmology based on general relativity with Galileo's physics. Dingle similarly claimed that the perfect cosmological principle has "precisely the same nature as perfectly circular orbits and immutable heavens" and that "it is largely identical with the Aristotelian principle of celestial regions."[38] It was and still is common to refer to the epicycles of ancient astronomy when scientists want to criticize a theory for being complicated and ad hoc.

---

[37] For an analysis of Einstein's attitude to and use of the history of science, see H. Kragh, *Einstein on the History and Nature of Science*, [in:] *The Way through Science and Philosophy*, eds. H. B. Andersen et al., College Publications, London 2006, pp. 99-118.

In other cases the references to history are not to concrete events or persons, but of the "history suggests" type where the record of some general idea in past science is used to evaluate the methodological basis of a modern theory. For example, string theory notoriously lacks connection to experiment and is, according to some critics, largely justified by the dubious idea that fundamental physics must be mathematically beautiful. One of the critics, Daniel Friedan, says: "History suggests that it is unwise to extrapolate to fundamental principles of nature from the mathematical forms used by theoretical physics in any particular epoch of its history, no matter how impressive their success. … Mathematical beauty in physics cannot be appreciated until after it has proved useful."[39]

Again, although the anthropic principle does not lead to precise predictions, it may be justified by referring to historical cases in which a theory has been highly successful in spite of its limited predictivity. The prime example of such a theory is Darwinian evolution, which is sometimes referred to in the debate over the standards to be used in fundamental physics and cosmology. "One is reminded of Darwin's theory, which is a powerful explanatory tool even though some question its predictive power," says Craig Hogan. "Anthropic arguments are vulnerable in the same way to

---

[38] H. Dingle, *Cosmology and Science, op. cit.*, p. 137.
[39] D. Friedan, *A Tentative Theory of Large Distance Physics*, "Journal of High Energy Physics" 2003, no. 10, 063.

'Just So' storytelling but may nevertheless form an important part of cosmological theory."[40]

One historical case that occurs surprisingly often in the universe-or-multiverse discussion is Kepler's geometrical model of the heliocentric universe as expounded in his *Mysterium Cosmographicum* from 1596. When multiverse proponents refer to Kepler's model, it is invariably as a negative exemplar, to illustrate that the universe is probably not uniquely described by the mathematical solutions to the equations of physics. According to Steven Weinberg, "We may just have to resign ourselves to a retreat, just as Newton had to give up Kepler's hope of a calculation of the relative sizes of planetary orbits from first principles."[41] Frank Wilczek uses the same case to argue for the same conclusion: "In the development of Copernican-Newtonian celestial mechanics, attractive a priori ideas about the perfect shape of planetary orbits (Ptolemy) and their origin in pure geometry (Kepler) had to be sacrificed."[42] On the other hand, Kepler may also be used as a positive exemplar (and Galileo as a negative exemplar), as Martin Rees does in his argument for the multiverse: "Kepler discovered that planets moved in ellipses, not circles. Galileo was upset by this. … The parallel is obvious. … A bias in favour of 'simple' cosmologies may be as short-sighted as was Galileo's infatuation with circles."[43]

---

[40] C. J. Hogan, *Why the Universe Is Just So*, "Reviews of Modern Physics" 2000, no. 72, pp. 1149-1161, on p. 1160.
[41] S. Weinberg, *Living in the Multiverse*, [in:] B. Carr, *Universe or Multiverse*, *op. cit.*, pp. 29-42, on p. 39.
[42] F. Wilczek, *Enlightenment, Knowledge, Ignorance, Temptation*, [in:] B. Carr, *Universe or Multiverse*, *op. cit.*, pp. 43-53, on p. 50.
[43] M. Rees, *Explaining the Universe*, [in:] J. Cornwell, *Explanation*, *op. cit.*, pp. 39-66, on p. 63.

My last example of the questionable use of history of science comes from Carr, who suggests that critics of the multiverse are on the wrong side of history. Throughout the history of cosmology, the universe has always been conceived as bigger and bigger, he claims, so why be satisfied with a single universe instead of a whole lot of them? Carr's argument may have some rhetorical force, but it is poor from both the perspective of history and from a logical point of view. At any rate, here it is:

> Throughout the history of science, the universe has always gotten bigger. We've gone from geocentric to galactocentric. Then in the 1920s there was this huge shift when we realized that our galaxy wasn't the universe. I just see this as one more step in the progression. Every time this expansion has occurred, the more conservative scientists have said, 'This isn't science.' This is the same process repeating itself.[44]

This is not the place for discussing the role of history of science in scientific or philosophical arguments, but it needs to be pointed out that in general one should be very cautious with reasoning based on historical analogies and extrapolations from historical trends. Historical arguments and analogies have a legitimate function in the evaluation of current science.[45] We cannot avoid being guided by the past, and it would be silly to disregard the historical record when thinking about the present and the future. On the other hand, such guidance should be based on historical insight and not, as is often the case, on arbitrary selections from a folk version of history.

---

[44] Quoted in T. Folger, *Science's Alternative to an Intelligent Creator: The Multiverse Theory*, "Discover Magazine" (online version) 2008. In fact, the universe has not "always gotten bigger." Kepler's universe was much smaller than Copernicus's, and Kant's universe of the 1750s was much bigger than the Milky Way universe a century later.

[45] L. Darden, *Viewing the History of Science as Compiled Hindsight*, "AI Magazine" 1987, no. 8:2, pp. 33-41. H. Kragh, *An Introduction to the Historiography of Science*, Cambridge University Press, Cambridge 1987, pp. 150-158.

Generally speaking, the history of science is so diverse and complex that it is very difficult to draw from it lessons of operational value for modern science. In 1956, in connection with the controversy over the steady state theory, Gold reflected on the lessons of history of science with regard to the methodology of cosmology and other sciences. He considered history to be an unreliable guide:

> Analogies drawn from the history of science are frequently claimed to be a guide [to progress] in science; but, as with forecasting the next game of roulette, the existence of the best analogy to the present is no guide whatever to the future. The most valuable lesson to be learned from the history of scientific progress is how misleading and strangling such analogies have been, and how success has come to those who ignored them.[46]

Of course, scientists should not ignore history. They can and should use the rich treasure of resources hidden in the history of science, but they must do it with proper caution and professional insight.

---

[46] T. Gold, *Cosmology*, "Vistas in Astronomy" 1956, no. 2, pp. 1721-1726, on p. 1722.